# Bose-Einstein condensation and independent production of pions

A.Bialas and K.Zalewski [*]

M.Smoluchowski Institute of Physics

Jagellonian University, Cracow[†]

August 13, 2018

## Abstract

The influence of the HBT effect on the momentum spectra of independently produced pions is studied using the method developed earlier for discussion of multiplicity distributions. It is shown that in this case all the spectra and multiparticle correlation functions are expressible in terms of one function of two momenta. It is also shown that at the critical point all pions are attracted into one quantum state and thus form a Bose-Einstein condensate.

1. Introduction

Several years ago, Pratt [1] realized that the well-established phenomenon of HBT correlations [2] can, under certain conditions, lead to Bose-Einstein condensation in multipion systems, which he called "pion laser". Since then the effect was investigated by several authors, including Pratt [3, 4] (see [5] for an exhaustive list of references). Recently [6], employing the density matrix formalism, we discussed multiplicity distributions of independently emitted identical particles for arbitrary shapes of particle spectra in momentum and

---

in configuration space. The approach to the condensation point and the conditions for reaching it were investigated. In the present paper we extend the method of [6] to momentum spectra of independently produced identical pions. The explicit formula for the generating functional is written down and the resulting momentum spectra are discussed.

Although the independent production mechanism is unlikely to be a realistic model of pion production, we feel that it is interesting to investigate its consequences.

**2.** The basic ideas of the HBT effect, as applied to processes of particle production, were explained in [6], using the approach developed in [7]. The net result is the formula for the momentum distribution of $n$ identical bosons

$$\Omega(q) == \frac{1}{n!} \sum_{P,P'} \rho^{(0)}(q_P, q_{P'}). \tag{1}$$

where $\rho^{(0)}(q, q')$ is the n-particle density matrix *calculated ignoring the identity of particles*[1] and normalized by the condition

$$Tr[\rho^{(0)}] = \int dq \rho^{(0)}(q, q) = 1. \tag{2}$$

The sum extends over all permutations $P$ and $P'$ of particle momenta $[q_1, ..., q_n] \equiv q$.

Consider now a system of $n$ particles emitted independently. If we ignore the identity of particles, independent emission implies that the density matrix factorizes

$$\rho^{(0)}(q, q') = \prod_{i=1}^{n} \rho^{(0)}(q_i, q'_i). \tag{3}$$

Introducing this into (1) we have

$$\Omega(q) = \frac{1}{n!} \sum_{P,P'} \prod_{i=1}^{n} \rho^{(0)}((q_P)_i, (q_{P'})_i). \tag{4}$$

The information contained in (4) is conveniently summarized in the form of the generating functional $\Phi[u]$ defined as

$$\Phi[u] = \frac{\sum_{n=0}^{\infty} P^{(0)}(n) W_n[u]}{\sum_{n=0}^{\infty} P^{(0)}(n) W_n[1]} \tag{5}$$

---

[1]All quantities calculated with the identity of particles ignored will be called *uncorrected* and denoted by a superscript $^{(0)}$.

with
$$W_n[u] = \int dq_1...dq_n \Omega(q_1,...,q_n) u(q_1)...u(q_n). \tag{6}$$

Here $P^{(0)}(n)$ is the *uncorrected* multiplicity distribution and $u(q)$ is an arbitrary real nonnegative function of $q$. For $u(q) = const \equiv z$ the generating functional reduces to the generating function of the multiplicity distribution, $\Phi(z)$ discussed in [6].

The inclusive and exclusive distributions of $n$ identical particles can be obtained by n-fold functional differentiation of the generating functional (5) with respect to $u(q)$ at $u(q) = 1$ and $u(q) = 0$, respectively. Similarly $n-$fold differentiation of the logarithm of $\Phi[u]$ at $u(q) = 1$ gives the inclusive correlation function of the n-th order.

To find the explicit expression for $\Phi[u]$ we observe that, given the formula (4) for $\Omega(q)$, for each permutation $P$ of the momenta $q_1, ..., q_n$, the integral on the right hand side of (6) factorizes into a product of contributions from all the cycles of P (as is well known, each permutation can be decomposed into cycles). Let us denote the contribution from a cycle of length $k$ by $C_k[u]$. We have

$$C_k[u] = \int d^3q_1...d^3q_k u(q_1)\rho^{(0)}(q_1,q_2) u(q_2)\rho^{(0)}(q_2,q_3)...u(q_k)\rho^{(0)}(q_k,q_1) \tag{7}$$

The rest of the calculation is just combinatorics.

We observe first that any two permutations which have identical partitions into cycles give equal contributions. Let us consider the set of all permutations with a given partition into cycles. Denoting by $n_k$ the number of occurrences of a cycle of length $k$ in the set of permutations considered, the contribution from all of them can be written as

$$W'_n[u] = \prod_{k=1}^{n} (C_k[u])^{n_k} \frac{n!}{(k!)^{n_k}} [(k-1)!]^{n_k} \frac{1}{n_k!} = n! \prod_{k=1}^{n} \frac{\left(\frac{C_k[u]}{k}\right)^{n_k}}{n_k!}. \tag{8}$$

In the first equality the first factor is the integral, the second is the number of partitions of the $n$ particles among the cycles, the third is the number of ways a cycle can be constructed from $k$ particles and the last one corrects for the permutations of whole cycles.

$W_n[u]$ is obtained by summing $W'_n[u]$ over partitions into cycles different from each other.

Until now we have considered a fixed multiplicity. As noted already in [6], the sum over multiplicities can be explicitly performed if the uncorrected multiplicity distribution $P^{(0)}(n)$ is poissonian (as required for independent emission)

$$P^{(0)}(n) = e^{-\nu}\frac{\nu^n}{n!}. \tag{9}$$

The result is an elegant formula for the generating functional (5):

$$\Phi[u] = \exp\left(\sum_{k=1}^{\infty} \nu^k \frac{C_k[u] - C_k[1]}{k}\right). \tag{10}$$

**3.** Let us now discuss the general properties of the particle distributions obtained from Eq.(10).

The main result is that all inclusive distributions can be expressed in terms of a single function $L(q, q')$, defined as

$$L(q, q') = \sum_{k=1}^{\infty} \nu^k [\rho^{(0)}]^k(q, q') \tag{11}$$

where

$$[\rho^{(0)}]^k(q, q') \equiv \int d^3q_2 ... d^3q_k \rho^{(0)}(q, q_2)\rho^{(0)}(q_2, q_3)...\rho^{(0)}(q_k, q') \tag{12}$$

and the $\nu$-dependence of $L$ is not written explicitly. The single particle distribution is given by

$$\omega(q) = L(q, q), \tag{13}$$

and the two-particle correlation function is

$$K_2(q_1, q_2) = L(q_1, q_2)L(q_2, q_1). \tag{14}$$

The general formula for the correlation functions reads

$$K_p(q_1, ...q_p) = L(q_1, q_2)L(q_2, q_3)....L(q_p, q_1)$$
$$+ permutations \ of \ (q_2, ....., q_p). \tag{15}$$

It is not difficult to verify that by integrating (15) over all momenta one recovers the formula for cumulants derived in [6].

These formulae represent a formidable constraint on the observed particle distributions: They basically say that all higher order correlations can be derived from the two particle correlation function. As they are valid for any model which assumes independent production, they were found by many authors in particular cases [9, 10, 11], for a full list, see [5]. It is fair to say, however, that - since the independent production model is not expected to be a precise description of high energy interactions one expects violations of these relations, at least to some extent. Recent work by Eggers et al. [12], points perhaps in this direction. The measurements of deviation of data from (13)-(15) is of great interest, as it may indicate what is the dominant intraparticle correlation.

Further discussion is greatly simplified if the matrix $\rho^{(0)}(q,q)$ is expressed in terms of its eigenvalues $\lambda_m$ and its eigenfunctions $\psi_m(q)$. We have[2]

$$\rho^{(0)}(q,q') = \sum_m \psi_m(q)\lambda_m \psi_m^*(q') \qquad (16)$$

and

$$[\rho^{(0)}]^k(q,q') = \sum_m \psi_m(q)(\lambda_m)^k \psi_m^*(q'). \qquad (17)$$

Substituting this into (11) and performing the sum over $k$ we have

$$L(q_1,q_2) = \sum_m \psi_m(q_1)\psi_m^*(q_2) \frac{\nu\lambda_m}{1-\nu\lambda_m}. \qquad (18)$$

It is clear from (18) that $L(q_1,q_2)$ and thus also all inclusive distributions become singular when $\nu\lambda_0 \to 1$ ($\lambda_0$ is the largest eigenvalue). In this limit, corresponding to Bose-Einstein condensation, $L(q,q')$ is dominated by the first term in the sum and we have

$$L(q_1,q_2) = \frac{\psi_0(q_1)\psi_0^*(q_2)}{1-\nu\lambda_0} + \tilde{L}(q_1,q_2), \qquad (19)$$

where $\tilde{L}(q_1,q_2)$ remains bounded for $\nu\lambda_0 \to 1$. Thus at the condensation point all the particles, except for a negligible fraction, are in the same state described by the eigenfunction $\psi_0(q)$. Away from the condensation point, of course, the whole sum in (18) must be carried out.

---

[2] We discuss here only the case of a discrete eigenvalue spectrum. The case of a continuous spectrum can be treated along the same lines [6].

Let us now consider the important example, when particles are emitted in a pure state, i.e.
$$\rho(q,q') = \psi(q)\psi^*(q'). \tag{20}$$
It follows from (7) that then
$$C_k[u] = (C_1[u])^k \tag{21}$$
and thus the generating functional becomes
$$\Phi[u] = \exp\left(\sum_{k=1}^{\infty} \frac{\nu^k((C_1[u])^k - 1)}{k}\right) = \frac{1-\nu}{1-\nu C_1[u]}. \tag{22}$$
One easily sees that this gives the momentum spectrum $|\psi(q)|^2$ independent of particle multiplicity and the geometric distribution of multiplicities [6].

**4.** Let us now consider the uncorrected single particle density matrix of gaussian form, discussed already in several papers by Pratt [1, 3, 4] and recently by other authors [5, 11]. To simplify the notation, we shall restrict ourselves to the one-dimensional problem. Generalization to the three-dimensional case is straightforward [6]. We thus have
$$\rho^{(0)}(q,q') = \left(\frac{1}{2\pi\Delta^2}\right)^{\frac{1}{2}} e^{-\frac{(q^+)^2}{2\Delta^2} - \frac{1}{2}R^2(q^-)^2}, \tag{23}$$
where
$$q^+ \equiv \frac{1}{2}(q+q'); \quad q^- \equiv q - q'. \tag{24}$$
As easily seen, $\Delta^2$ is the average value of the square of the particle momentum, and $R^2$ is the average value of the square of the space coordinate of the particle emission point. As is clear from the context, both $\Delta$ and $R$ refer to the *uncorrected* distributions. The uncertainty principle implies that
$$R\Delta \geq \frac{1}{2}. \tag{25}$$

As explained in [6], the eigenfunctions of the density matrix (23) are of the form
$$\psi_m(q) = a_m e^{-\frac{1}{2}\frac{R}{\Delta}q^2} H_m(\sqrt{\frac{R}{\Delta}}q), \tag{26}$$

where $H_m(q)$ is the Hermite polynomial of order $m$ and $a_m$ is the normalizing factor, given e.g. in [13].

The corresponding eigenvalues are

$$\lambda_m = \lambda_0(1-\lambda_0)^m, \quad m = 0, 1, ..., \tag{27}$$

where

$$\lambda_0 = \frac{2}{(1+2\Delta R)} \leq 1 \tag{28}$$

is the greatest one.

Eqs. (26) and (27) can be now used for the explicit calculation of $L(q, q')$ [c.f. (18)] which, in turn, determines all particle distributions, as explained in the previous section. This is actually easier than it looks because the difficult sum over oscillating Hermite polynomials can be replaced by a sum over Gausssians with positive coefficients. To see this, we rewrite the formula (18) as a double sum

$$L(q, q') = \sum_m \psi_m(q)\psi_m^*(q') \sum_{k=1}^{\infty} (\nu\lambda_m)^k. \tag{29}$$

Introducing (27) and reversing the order of summation we obtain

$$L(q, q') = \sum_{k=1}^{\infty} \frac{(\nu\lambda_0)^k}{1-(1-\lambda_0)^k}\hat{\rho}_k(q, q'), \tag{30}$$

where

$$\hat{\rho}_k(q, q') = \left(\frac{1}{2\pi\hat{\Delta}_k^2}\right)^{\frac{1}{2}} e^{-\frac{(q^+)^2}{2\hat{\Delta}_k^2} - \frac{1}{2}\hat{R}_k^2(q^-)^2}, \tag{31}$$

and $\hat{\Delta}_k$ and $\hat{R}_k$ are determined from the equations

$$\frac{\hat{R}_k}{\hat{\Delta}_k} = \frac{R}{\Delta}; \quad \hat{R}_k\hat{\Delta}_k = \frac{1}{2}\frac{1+\omega_k}{1-\omega_k}; \quad \omega_k = \left(\frac{2R\Delta-1}{2R\Delta+1}\right)^k. \tag{32}$$

The important lesson from this exercise is that even when the uncorrected distribution is described by a simple Gaussian, the resulting particle spectra are fairly complicated superpositions of an infinite number of Gaussians with varying width.

It follows from (32) that $\hat{R}_k < R$ and $\hat{\Delta}_k < \Delta$ for all $k > 1$. Consequently, the observed distributions are always *narrower* than the assumed uncorrected ones.

Another interesting quantity is the average value $< (q^-)^2 >$ calculated from the two-particle correlation function, i.e.,

$$< (q^-)^2 > = \frac{1}{K_2} \int dq^+ K_2(q, q')(q^-)^2 \qquad (33)$$

where the cumulant $K_2$ is the integral of $K_2(q, q')$. In the standard analysis of the data, $2 < (q^-)^2 >$ is usually interpreted as an inverse of the average squared radius $R_{eff}$ of the particle emission region. The ratio $R_{eff}^2/R^2$ calculated from (33) is plotted in Figure 1 for various values of $R\Delta$. One sees that, at a fixed $R\Delta$, $R_{eff}^2$ decreases from $R^2$ to $R^2/2R\Delta$ when $<n>$ varies from 0 to $\infty$. Thus we conclude that, even fairly far from the critical point, the apparent size of the system $R_{eff}^2$ (as determined from the two-particle correlation function) has little to do with the actual size of the system, given by $R^2$. One sees also that, even at the fixed particle phase-space density $<n>/R\Delta$, the effect substantially increases with increasing $R\Delta$.

**5.** Our conclusions can be summarized as follows.

(i) It has been shown that the formulation of the problem of HBT interference in terms of the density matrix provides an effective tool for discussion not only of particle multiplicities [6] but also of the momentum spectra.

(ii) For independent particle production, the effects of HBT symmetrization are expressed in terms of one function of two momenta. Consequently, all multiparticle correlations are expressible in terms of the single particle distribution and the two-particle correlation function. It would be very interesting to verify if this general relation holds – at least approximately – in the data.

(iii) The symmetrized distributions reveal the existence of a critical point corresponding to the Bose-Einstein condensation, when almost all the particles fall into one quantum state, corresponding to the lowest eigenvalue of the (uncorrected) density matrix.

(iv) Due to symmetrization, the width of the momentum spectrum decreases and the width of the two-particle correlation function increases. This effect becomes stronger, when the system approaches criticality. Close to the critical point the determination of the size of the emission region from the width of the correlation function is no longer possible.

(v) For a Gaussian uncorrected density matrix, the corrected momentum distribution is Gaussian both in the low density and in the high density limit, but non-Gaussian in the intermediate region. At low density the parameters $\Delta$ and $R$ are determined by the widths of the momentum distribution and of the correlation function, respectively. At high density both these widths depend on the ratio $\frac{R}{\Delta}$ only.

**Acknowledgements**

We would like to thank P.Bialas, K. Fia?lkowski, J.Pisut and A.Staruszkiewicz for very useful comments and discussions. This work was supported in part by the KBN Grant 2 P03B 086 14.

**Figure captions**

Figure 1. $R_{eff}/R^2$ plotted versus $<n>$ for different values of $R\Delta$. The marked points correspond to a fixed value of the particle phase-space density $<n>/R\Delta = 8$.